\documentclass[conference]{IEEEtran}
\IEEEoverridecommandlockouts

\usepackage{cite}

\usepackage{amsmath}
\usepackage{amssymb}
\usepackage{mathdots}
\usepackage{algorithmic}
\usepackage{algorithm}
\usepackage{array}
\usepackage{multirow}
\usepackage{tikz}
\usepackage{pgfplots}
\pgfplotsset{compat=newest}
\pgfplotsset{legend style={rounded corners=2pt,nodes=right}}
\DeclareMathOperator{\fieldR}{\mathbb{R}}
\DeclareMathOperator{\fieldC}{\mathbb{C}}

\DeclareMathOperator{\T}{\operatorname{T}}
\DeclareMathOperator{\Herm}{\operatorname{H}}

\newcommand{\herm}[1]{#1^{\mathrm{H}}}
\newcommand{\ve}[1]{\boldsymbol{#1}}

\newcommand{\e}{\mathrm{e}}
\newcommand{\E}{\mathrm{E}}
\begin{document}
\newlength{\figurewidth}
\newlength{\figureheight}

\title{Analog Transmit Signal Optimization for Undersampled Delay-Doppler Estimation}

\author{\IEEEauthorblockN{Andreas Lenz\IEEEauthorrefmark{1},
Manuel S. Stein\IEEEauthorrefmark{2},
A. Lee Swindlehurst\IEEEauthorrefmark{3}}
\IEEEauthorblockA{\IEEEauthorrefmark{1}Institute for Communications Engineering, Technische Universit\"at M\"unchen, Germany}
\IEEEauthorblockA{\IEEEauthorrefmark{2}Mathematics Department, Vrije Universiteit Brussel, Belgium}
\IEEEauthorblockA{\IEEEauthorrefmark{3}Henry Samueli School of Engineering, University of California, Irvine, USA}
E-Mail: andreas.lenz@mytum.de, manuel.stein@vub.ac.be, swindle@uci.edu
\thanks{This work was supported by the EIKON e.V., the Heinrich and Lotte M\"{u}hlfenzl Foundation and the Institute for Advanced Study (IAS), Technische Universit\"{a}t M\"{u}nchen (TUM), with funds from the German Excellence Initiative and the European Union's Seventh Framework Program (FP7) under grant agreement no.~291763. This work was also supported by the German Academic Exchange Service (DAAD) with funds from the German Federal Ministry of Education and Research (BMBF) and the People Program (Marie Curie Actions) of the European Union's Seventh Framework Program (FP7) under REA grant agreement no.~605728 (P.R.I.M.E. - Postdoctoral Researchers International Mobility Experience).}
}

\maketitle

\begin{abstract}
In this work, the optimization of the analog transmit waveform for joint delay-Doppler estimation under sub-Nyquist conditions is considered. Based on the Bayesian Cram\'er-Rao lower bound (BCRLB), we derive an estimation theoretic design rule for the Fourier coefficients of the analog transmit signal when violating the sampling theorem at the receiver through a wide analog pre-filtering bandwidth. For a wireless delay-Doppler channel, we obtain a system optimization problem which can be solved in compact form by using an Eigenvalue decomposition. The presented approach enables one to explore the Pareto region spanned by the optimized analog waveforms. Furthermore, we demonstrate how the framework can be used to reduce the sampling rate at the receiver while maintaining high estimation accuracy. Finally, we verify the practical impact by Monte-Carlo simulations of a channel estimation algorithm.
\end{abstract}

\setlength{\figurewidth}{0.4\textwidth}
\setlength{\figureheight}{5.2cm}

\IEEEpeerreviewmaketitle
\section{Introduction}%
\IEEEPARstart{C}{hannel} parameter estimation enjoys significant attention in the signal processing literature and is key to applications, such as radar and mobile communication. Radar systems use knowledge of the delay-Doppler shift to precisely determine the position and velocity of a target object, while in wireless communication channel estimation is required for beamforming techniques and rate adaptation.

In signal processing systems, the  prevailing design paradigm for the bandwidth of the transmit and receive filter is compliance with the well-known sampling theorem, requiring a sufficiently high receive sampling rate. While this guarantees perfect signal reconstruction from the receive data, it stands in contrast to results from estimation theory, where high bandwidths can be beneficial for parameter estimation, see e.g. \cite{Sadler06}. When the receive system is designed such that it satisfies the sampling theorem, i.e., the analog pre-filter bandlimits the sensor signal to the analog-to-digital conversion rate, the achievable sampling rate $f_s$ at the receiver restricts the bandwidth $B$ of the transmitter and therefore the overall system performance. Since the sampling rate forms a bottleneck with respect to power resources and hardware limitations \cite{Verhelst15}, it is necessary to find a trade-off between high performance and low complexity. Therefore we discuss how to design the transmit signal for delay-Doppler estimation without the commonly used restriction from the sampling theorem.

Delay-Doppler estimation has been discussed for decades in the signal processing community \cite{Jakobsson98, Jakobsson98_2, Friedlander84}. In \cite{Jakobsson98} a subspace based algorithm for the estimation of multi-path delay-Doppler shifts is proposed and it is shown how the dimensionality of the maximum likelihood (ML) estimator can be reduced by a factor of two. In \cite{Jakobsson98_2} a time-domain procedure for estimation of delay-Doppler shifts and direction of arrival (DOA) is considered. Using prolate spheroidal wave (PSW) functions, the favorable transmit signal design with respect to time-delay accuracy is discussed in \cite{Antreich11}, while \cite{Jin95} considers such a technique for joint delay-Doppler estimation. Recent results show that for wide-band transmit signals, analog receive filter bandwidths which lead to violation of the sampling theorem can provide performance gains \cite{Stein14,Lenz15}. Further, in \cite{Khayambashi14} the optimization of receive filters in a compressed sensing framework has been investigated and improvements with respect to matched filtering have been illustrated. 

Here we consider optimization of the transmit signal while the receiver samples at a rate $f_s$ smaller than the Nyquist rate $B$. After introducing the system model for a single-input single-output (SISO) delay-Doppler channel, we derive a compact formulation of the transmit signal optimization problem in the frequency domain. We show how to solve the transmitter design problem for $B>f_s$ by an Eigenvalue decomposition. The potential Pareto-optimal region is visualized by optimizing the transmit waveform for different settings and comparing the results to conventional signal designs. We conclude the discussion with a performance verification via Monte-Carlo simulations of a channel estimation algorithm.
\section{System Model}
Consider the propagation of an analog, $T_0$-periodic pilot signal $\breve{x}(t) \in \mathbb{C}$ through a wireless delay-Doppler channel. The baseband signal at the receiver, which is perturbed by additive white Gaussian noise (AWGN) $\breve{\eta}(t) \in \mathbb{C}$ with constant power spectral density $N_0$, can be denoted as
\begin{equation}\label{signal:receiver:unfiltered}
\breve{y}(t) = \gamma \breve{x}(t-\tau) \e^{\mathrm{j} 2\pi \nu t} + \breve{\eta}(t)
\end{equation}
with channel coefficient $\gamma \in \mathbb{C}$, time-delay $\tau \in \mathbb{R}$ and Doppler shift $\nu \in \mathbb{R}$. The signal $\breve{y}(t)\in\mathbb{C}$ is filtered by a linear receive filter $h(t) \in \mathbb{C}$, such that the final analog receive signal
\begin{align}\label{eq:Filtered_Receive_Signal}
y(t) &= \big(\gamma \breve{x}(t-\tau) \e^{\mathrm{j} 2\pi \nu t} + \breve{\eta}(t) \big)*h(t)\notag\\
&= v(t;\boldsymbol{\theta}) + \eta(t)
\end{align}
is obtained, where $\boldsymbol{\theta} = \begin{pmatrix}\tau &\nu\end{pmatrix}^\mathrm{T} \in \fieldR^2$
denotes the unknown, random channel parameters. For the duration $T_0$, the signal $y(t)\in\mathbb{C}$ is sampled in intervals of $T_s = \frac{1}{f_s}$, resulting in an even number of $N = \frac{T_0}{T_s} \in 2\mathbb{N}$ samples
\begin{equation}\label{eq:Sampled_Receive_Signal}
\boldsymbol{y} = \boldsymbol{v} (\boldsymbol{\theta}) + \boldsymbol{\eta},
\end{equation}
with the receive vectors $\ve{y}, \ve{v}(\ve{\theta}), \ve{\eta} \in \mathbb{C}^N$ defined as
\begin{align}
[\ve{y}]_i &= y\left(\left(i-\frac{N}{2}-1\right)T_s\right), \\
[\ve{v}(\ve{\theta})]_i &= v\left(\left(i-\frac{N}{2}-1\right)T_s, \ve{\theta}\right), \\
[\ve{\eta}]_i &= \eta\left(\left(i-\frac{N}{2}-1\right)T_s\right).
\label{eq:Sampled_Signals}
\end{align}
We use positive integers as indices for vectors and matrices and thus $i\in \{1, 2, \dots, N\}$. The noise samples $\boldsymbol{\eta}$ in \eqref{eq:Sampled_Receive_Signal} follow a zero-mean Gaussian distribution with covariance matrix
\begin{equation}\label{eq:noise:covariance:matrix}
\boldsymbol{R}_{\boldsymbol{\eta}} = \E_{\boldsymbol{\eta}}[\boldsymbol{\eta}\boldsymbol{\eta}^{\Herm}] \in \mathbb{C}^{N \times N}.
\end{equation}
Note that $\ve{R}_{\ve{\eta}}$ depends on the receive filter $h(t)$ and the sampling rate $f_s$ and thus is not necessarily a scaled identity matrix. The unknown parameters $\ve{\theta}$ are considered to be Gaussian distributed $p(\ve{\theta}) \sim \mathcal{N}(\ve{0}, \ve{R}_{\ve{\theta}})$ with known covariance
\begin{equation}
\ve{R}_{\ve{\theta}} = \begin{pmatrix}
\sigma_\tau^2 & 0 \\ 0 & \sigma_\nu^2
\end{pmatrix}.
\end{equation}
Here we assume that the channel $\gamma$ is known at the receiver, which simplifies the formulation of the transmit signal optimization problem. However, when testing the optimized waveforms for a practical scenario in the last section we will treat $\gamma$ to be a deterministic unknown. For the derivation, we first assume a fixed sampling rate $f_s$ at the receiver while the periodic transmit signal $\breve{x}(t)$ is band-limited with two-sided bandwidth $B$. Then we consider the case of a variable rate $f_s$. In contrast to the sampling theorem assumption $B \leq f_s$, in our setup we allow $B > f_s$. Note that at the receiver, we always use an ideal low-pass filter $h(t)$ featuring the same bandwidth $B$ as the transmit signal.

\section{Channel Estimation Problem}
Under the assumption that $\gamma$ is known, the task of the receiver is to infer the unknown channel parameters $\boldsymbol{\theta}$ based on the digital receive data $\boldsymbol{y}$ using an appropriate channel estimation algorithm $\boldsymbol{\hat{\theta}} (\boldsymbol{y})$.  The mean squared error (MSE) of the estimator $\boldsymbol{\hat{\theta}} (\boldsymbol{y})$ is defined as
\begin{equation}\label{eq:MSE:matrix}
\boldsymbol{R}_{\ve{\epsilon}} = \E_{\boldsymbol{y}, \boldsymbol{\theta}} \Big[\big(  \boldsymbol{\hat\theta}(\boldsymbol{y}) - \boldsymbol{\theta} \big) \big( \boldsymbol{\hat\theta}(\boldsymbol{y}) - \boldsymbol{\theta} \big)^{\T}\Big].
\end{equation}
A fundamental limit for the estimation accuracy \eqref{eq:MSE:matrix} is the \textit{Bayesian Cram\'{e}r-Rao lower bound} (BCRLB) \cite[p. 5]{Trees07}
\begin{equation}\label{eq:BCRLB}
\boldsymbol{R}_{\ve{\epsilon}} \succeq \ve{J}_B^{-1},
\end{equation}
where $\ve{J}_B$ is the \textit{Bayesian information matrix} (BIM)
\begin{equation}\label{eq:BIM}
\ve{J}_B = \ve{J}_D + \ve{J}_P.
\end{equation}
The first summand of the BIM \eqref{eq:BIM} represents the \textit{expected Fisher information matrix} (EFIM) 
\begin{equation}\label{eq:EFIM}
\boldsymbol{J}_D  = \E_{\boldsymbol{\theta}}\big[\boldsymbol{J}_F(\boldsymbol{\theta})\big],
\end{equation}
with the \textit{Fisher information matrix} (FIM) exhibiting entries
\begin{equation}\label{eq:FIM:entries}
[\boldsymbol{J}_F(\boldsymbol{\theta})]_{ij} = - \E_{\boldsymbol{y}|\boldsymbol{\theta}} \Bigg[ \frac{\partial^2 \ln p (\boldsymbol{y}| \boldsymbol{\theta})}{\partial [\boldsymbol{\theta}]_i \partial [\boldsymbol{\theta}]_j} \Bigg].
\end{equation}
For the signal model \eqref{eq:Sampled_Receive_Signal}, the FIM entries \eqref{eq:FIM:entries} are
\begin{equation}\label{eq:FI_Entries_General}
[\boldsymbol{J}_F(\boldsymbol{\theta})]_{ij} = 2 \mathrm{Re} \left\{  \herm{\left( \frac{\partial \boldsymbol{v} (\boldsymbol{\theta})}{\partial [\boldsymbol{\theta}]_i} \right)} \boldsymbol{R}_{\boldsymbol{\eta}}^{-1} \left( \frac{\partial \boldsymbol{v} (\boldsymbol{\theta})}{\partial [\boldsymbol{\theta}]_j} \right) \right\}. 
\end{equation}
The second summand in \eqref{eq:BIM} denotes the \textit{prior information matrix} (PIM) $\boldsymbol{J}_P$ with entries
\begin{equation}
[\boldsymbol{J}_P]_{ij} =  - \E_{\boldsymbol{\theta}}\Bigg[ \frac{\partial^2 \ln p(\boldsymbol{\theta})}{\partial [\boldsymbol{\theta}]_i \partial [\boldsymbol{\theta}]_j} \Bigg].
\end{equation}
\section{Transmitter Optimization Problem}
The design problem of finding a transmit signal $\breve{x}^\star(t)$ that minimizes the MSE \eqref{eq:MSE:matrix} of the estimation algorithm $\boldsymbol{\hat{\theta}} (\boldsymbol{y})$ under a particular positive semi-definite weighting $\ve{M} \in \mathbb{R}^{2\times 2}$, subject to a transmit power constraint $P$, can be phrased as
\begin{equation}\label{eq:optimization:problem:initial}
\breve{x}^\star(t) = \underset{\breve{x}(t)}{\arg \min}\,\mathrm{tr} (\boldsymbol{M} \boldsymbol{R}_{\ve{\epsilon}}),\,\mathrm{s.t.}\,\frac{1}{T_0} \int_{T_0} |\breve{x}(t)|^2 \mathrm{d}t \leq P.
\end{equation}
Although the BCRLB \eqref{eq:BCRLB} can be achieved with equality only under special conditions \cite[p. 5]{Trees07}, it closely characterizes the estimation performance trend (see Sec. \ref{ss:Simulation_Results}). It is hence possible to formulate \eqref{eq:optimization:problem:initial} based on the BIM \eqref{eq:BIM}
\begin{equation}\label{eq:Opt_Prob}
\breve{x}^\star(t) = \underset{\breve{x}(t)}{\arg \min}\,\mathrm{tr} (\boldsymbol{M} \boldsymbol{J}_B^{-1}),\,\mathrm{s.t.}\frac{1}{T_0} \int_{T_0} |\breve{x}(t)|^2 \mathrm{d}t \leq P.
\end{equation}
In order to avoid optimization with respect to $\boldsymbol{J}_B^{-1}$ in \eqref{eq:Opt_Prob}, we consider an alternative maximization problem
\begin{equation} \label{eq:Opt_Prob_Alt}
\breve{x}^\star(t) = \underset{\breve{x}(t)}{\arg \max}\,\mathrm{tr} (\boldsymbol{M'} \boldsymbol{J}_{B}),\,\,\,\mathrm{s.t.}\,\,\,\,\frac{1}{T_0} \int_{T_0} |\breve{x}(t)|^2 \mathrm{d}t \leq P.
\end{equation}
It can been shown that if $\breve{x}^\star(t)$ is a solution of the maximization problem \eqref{eq:Opt_Prob_Alt} with $\boldsymbol{M'}$, there exists a weighting matrix $\boldsymbol{M}$ (not necessarily equal to $\ve{M}'$) for which the original minimization problem \eqref{eq:Opt_Prob} has the same solution $\breve{x}^\star(t)$ \cite{Stein14_2}. Since $\ve{J}_P$ is independent of $\breve{x}(t)$, \eqref{eq:Opt_Prob_Alt} then simplifies to
\begin{equation} \label{eq:Opt_Prob_Alt_Simplified}
\begin{aligned}
\breve{x}^\star(t) = \underset{\breve{x}(t)}{\arg \max}\,\mathrm{tr} (\boldsymbol{M'} \boldsymbol{J}_{D}), \,\,\,\mathrm{s.t.}\,\,\,\,\frac{1}{T_0} \int_{T_0} |\breve{x}(t)|^2 \mathrm{d}t \leq P.
\end{aligned}
\end{equation}
\section{Estimation Theoretic Performance Measure}
\label{section:performance:measure:delay:doppler}
Solving the optimization problem \eqref{eq:Opt_Prob_Alt} requires an analytical characterization of the EFIM \eqref{eq:EFIM}. A frequency-domain representation enables a compact notation of the receive signal model \cite{Lenz15} and thus provides further insights on the FIM entries \eqref{eq:FI_Entries_General}. Note that a frequency-domain approach naturally embodies the bandwidth restriction required in practice by limiting the number of Fourier coefficients.
\subsection{Signal Frequency Domain Representation}
Due to periodicity, the transmit waveform $\breve{x}(t)$ can be represented by its Fourier series
\begin{equation}
\breve{x}(t) = \sum_{k=-\frac{K}{2}}^{\frac{K}{2}-1} X_k \e^{\mathrm{j} k \omega_0 t}, \label{eq:Periodic_Transmit_Signal}
\end{equation}
where $\omega_0=\frac{2\pi}{T_0} = 2\pi f_0$ and $K= \lceil \frac{2\pi B}{\omega_0} \rceil \in 2\mathbb{N}$ is the total number of harmonics. $X_k$ denotes the $k$-th Fourier coefficient of the transmit signal. Inserting expression (\ref{eq:Periodic_Transmit_Signal}) into (\ref{eq:Filtered_Receive_Signal}) and applying the filtering operation in (\ref{eq:Filtered_Receive_Signal}), we obtain
\begin{align}\label{transmit:signal:receiver:filtered:exact:fdomain}
v(t;\ve{\theta}) &= \gamma \sum_{k=-\frac{K}{2}}^{\frac{K}{2}-1} X_k \Big(\e^{\mathrm{j}k\omega_0(t-\tau)} \e^{\mathrm{j} 2 \pi \nu t}\Big) * h(t) \nonumber \\
&= \gamma \e^{\mathrm{j}2\pi \nu t} \sum_{k=-\frac{K}{2}}^{\frac{K}{2}-1} \e^{\mathrm{j} k\omega_0 t} \e^{-\mathrm{j}k\omega_0 \tau} H(k\omega_0 + 2\pi \nu) X_k,
\end{align}
where $H(\omega)$ is the Fourier transform of the receive filter $h(t)$.
Evaluating $v(t;\ve{\theta})$ at instants $nT_s, n =-\frac{N}{2}, \dots, \frac{N}{2}-1$ yields
\begin{align}\label{eq:transmit:signal:sampled:fdomain}
v(nT_s;\ve{\theta}) =& \gamma \sum_{k=-\frac{K}{2}}^{\frac{K}{2}-1} \e^{\mathrm{j} 2\pi \nu n T_s} \e^{\mathrm{j}2\pi\frac{kn}{N}} \e^{-\mathrm{j}k\omega_0 \tau} H(k\omega_0 + 2\pi \nu) X_k \nonumber \\
=& \sum_{k=-\frac{K}{2}}^{\frac{K}{2}-1} [\ve{C}(\ve{\theta})]_{n+\frac{N}{2}+1,k+\frac{K}{2}+1} X_k,
\end{align}
with the channel matrix $\ve{C} (\ve{\theta}) \in \fieldC^{N \times K}$, defined by
\begin{equation} \label{eq:channel:matrix}
\ve{C} (\ve{\theta}) = \gamma \sqrt{N} \ve{D}(\nu) \herm{\ve{W}} \ve{T}(\tau) \ve{H}(\nu).
\end{equation}
The indices of $\ve{C}(\ve{\theta})$ in \eqref{eq:transmit:signal:sampled:fdomain} stem from the fact that we use positive integers as indices for vectors and matrices. Here $\ve{D}(\nu) \in \mathbb{C}^{N \times N}$ stands for a diagonal matrix
\begin{align}
[\ve{D}(\nu)]_{ii} &= \e^{\mathrm{j}2\pi \left(i-\frac{N}{2}-1\right) \nu T_s},
\end{align}
which represents the Doppler frequency-shift. Further $\ve{W} \in \mathbb{C}^{K \times N}$ is a tall discrete Fourier transform (DFT) matrix
\begin{align}
\left[\ve{W}\right]_{ij} &= \frac{1}{\sqrt{N}} \e^{-\mathrm{j}2\pi \frac{\left( i-\frac{K}{2} -1 \right) \left(j-\frac{N}{2}-1 \right) }{N}},
\end{align} 
and $\ve{T}(\tau) \in \mathbb{C}^{K \times K}$ denotes the diagonal time-delay matrix
\begin{align}\label{eq:tau:matrix:entries}
[\ve{T}(\tau)]_{ii} &= \e^{-\mathrm{j} \left(i-\frac{K}{2}-1\right) \omega_0 \tau}.
\end{align}
The diagonal matrix $\ve{H}(\nu)\in \mathbb{C}^{K \times K}$ in \eqref{eq:channel:matrix} characterizes the frequency shifted receive filter spectrum and has elements
\begin{align}\label{eq:h_d_tilde}
[\ve{H}(\nu)]_{ii} &= H\Bigg( \bigg(i-\frac{K}{2}-1\bigg)\omega_0 + 2\pi \nu \Bigg).
\end{align}
Note that the channel matrix \eqref{eq:channel:matrix} describes the propagation of $\ve{\tilde{x}}$ through the channel and its transformation from the spectral to the time domain. Further note that the aliasing effect due to bandwidths $B$ higher than the sampling frequency $f_s$ is automatically included by the wide IDFT matrix $\herm{\ve{W}}$.

Stacking the entries of $v(nT_s; \ve{\theta})$ \eqref{eq:transmit:signal:sampled:fdomain} into one vector yields 
\begin{equation} \label{eq:transmit:signal:sampled:filter:fdomain}
\ve{v}(\ve{\theta}) = \ve{C}(\ve{\theta}) \ve{\tilde{x}},
\end{equation}
with the transmit filter spectrum vector $\ve{\tilde{x}}\in \mathbb{C}^{K}$ formed by the Fourier coefficients
\begin{align}\label{frequency:representation:transmit:filter}
[\ve{\tilde{x}}]_i &= X_{i-\frac{K}{2}-1}.
\end{align}
\subsection{Fisher Information of the Delay-Doppler Channel}
In order to compute the FIM elements \eqref{eq:FI_Entries_General}, it is necessary to compute the derivatives of $\ve{v}(\ve{\theta})$ with respect to the parameters $\ve{\theta}$. Using the frequency domain representation \eqref{eq:transmit:signal:sampled:filter:fdomain}, we obtain
\begin{equation}
	\frac{\partial}{\partial [\ve{\theta}]_i} \ve{v}(\ve{\theta}) = \frac{\partial \ve{C}(\ve{\theta})}{\partial [\ve{\theta}]_i} \ve{\tilde{x}}.
\end{equation}
The derivatives of the channel matrix are
\begin{align}
\frac{\partial \ve{C}(\ve{\theta})}{\partial \tau} =& \gamma\sqrt{N}  \ve{D}(\nu) \ve{W}^\mathrm{H} \ve{\partial T}(\tau) \ve{H}(\nu) ,
\label{eq:D_V_D_Tau} \\
\frac{\partial \ve{C}(\ve{\theta})}{\partial \nu} =& \gamma \sqrt{N} \Big(\ve{\partial D}(\nu) \ve{W}^\mathrm{H} \ve{T}(\tau) \ve{H}(\nu) + \nonumber \\  & \qquad \qquad \,\, \ve{D}(\nu) \ve{W}^\mathrm{H} \ve{T}(\tau) \ve{\partial H}(\nu) \Big), \label{eq:D_V_D_Nu}
\end{align}
with the partial derivatives
\begin{align}
[\ve{\partial D}(\nu)]_{ii} &= \mathrm{j}2\pi \bigg(i-\frac{N}{2}-1\bigg) T_s \e^{\mathrm{j}2\pi \left(i-\frac{N}{2}-1\right) \nu T_s}, \\
[\ve{ \partial T}(\tau)]_{ii} &= -\mathrm{j} \left(i-\frac{K}{2}-1 \right) \omega_0 \e^{-\mathrm{j} \left(i-\frac{K}{2}-1 \right) \omega_0 \tau},\\
[\ve{\partial H}(\nu)]_{ii} &= \frac{\partial}{\partial \nu} H\left( \bigg(i-\frac{K}{2}-1\bigg)\omega_0 + 2\pi \nu \right).
\end{align}
Inserting \eqref{eq:D_V_D_Tau} and \eqref{eq:D_V_D_Nu} into \eqref{eq:FI_Entries_General}, the FIM entries can be expressed as quadratic terms
\begin{align}
[\ve{J}_F(\ve{\theta})]_{ij} &= 2 \mathrm{Re} \left\{ \herm{\ve{\tilde{x}}} \frac{\partial \herm{\ve{C}}(\ve{\theta})}{\partial [\ve{\theta}]_i} \ve{R}_{\ve{\eta}}^{-1} \frac{\partial \ve{C}(\ve{\theta})}{\partial [\ve{\theta}]_j} \ve{\tilde{x}} \right\}.
\end{align}
The elements of the expected Fisher information matrix (EFIM) \eqref{eq:EFIM} are then obtained by 
\begin{align}\label{compact:expression:EFIM:entries}
[\ve{J}_D]_{ij} &= 2 \mathrm{Re} \left\{ \herm{\ve{\tilde{x}}}  \E_{\ve{\theta}} \left[\frac{\partial \herm{\ve{C}}(\ve{\theta})}{\partial [\ve{\theta}]_i} \ve{R}_{\ve{\eta}}^{-1} \frac{\partial \ve{C}(\ve{\theta})}{\partial [\ve{\theta}]_j} \right] \ve{\tilde{x}} \right\} \nonumber \\
&= \herm{\ve{\tilde{x}}} \left(\ve{\Gamma}_{ij} + \ve{\Gamma}_{ji}\right) \ve{\tilde{x}},
\end{align}
with the channel sensitivity matrix $\ve{\Gamma}_{ij} \in \mathbb{C}^{K \times K}$
\begin{equation}\label{eq:channel:sensitivity:matrix}
\ve{\Gamma}_{ij} = \E_{\ve{\theta}} \left[\frac{\partial \herm{\ve{C}}(\ve{\theta})}{\partial [\ve{\theta}]_i} \ve{R}_{\ve{\eta}}^{-1} \frac{\partial \ve{C}(\ve{\theta})}{\partial [\ve{\theta}]_j} \right] .
\end{equation}
\section{Transmit Signal Optimization}
\label{sec:optimization:algorithm}
In the following we solve the transceiver design problem \eqref{eq:Opt_Prob_Alt} using the EFIM expressions \eqref{compact:expression:EFIM:entries}. With the frequency domain representation \eqref{frequency:representation:transmit:filter} of the transmit signal, the optimization problem \eqref{eq:Opt_Prob_Alt} becomes a maximization with respect to the transmit Fourier coefficients $\ve{\tilde{x}}$
\begin{equation}\label{eq:Opt_Approx:frequency:representation}
\ve{\tilde{x}}^\star = \underset{\ve{\tilde{x}}}{\arg \max}\quad \mathrm{tr} \left(\boldsymbol{M}' \ve{J}_D\right) \quad \mathrm{s.t.}\quad\herm{\ve{\tilde{x}}} \ve{\tilde{x}} \leq P.
\end{equation}
Expanding the trace operation, the objective function becomes
\begin{equation}\label{definition:objective:function:iterative:algorithm}
\mathrm{tr} \left(\boldsymbol{M}' \ve{J}_D\right) = \sum_{i=1}^{2} \sum_{j=1}^{2} [\boldsymbol{M}']_{ji} [\ve{J}_D]_{ij}
= \herm{\ve{\tilde{x}}}\ve{\Gamma}\ve{\tilde{x}},
\end{equation}
with the weighted channel sensitivity matrix
\begin{equation}
\ve{\Gamma} = \sum_{i=1}^{2} \sum_{j=1}^{2} [\boldsymbol{M}']_{ji} \left(\ve{\Gamma}_{ij} + \ve{\Gamma}_{ji}\right).
\end{equation}
The solution to the problem \eqref{eq:Opt_Approx:frequency:representation} is the Eigenvector $\boldsymbol{\gamma}_1$ of the matrix $\boldsymbol{\Gamma}$ corresponding to its largest Eigenvalue.
\section{Results}
\label{sec:simulation:results}
There exists a trade-off between the estimation of delay and Doppler-shift. By solving the optimization problem \eqref{eq:Opt_Approx:frequency:representation} for all positive semi-definite weightings $\ve{M}'$, we are able to approximate the Pareto-optimal region. This region is characterized by the set of transmit waveforms for which the estimation of one parameter cannot be improved by changing the transmit signal without reducing the accuracy of the other parameter.

For visualization, we define the relative measures
\begin{equation}\label{def:relative:performance}
\chi_{\tau/\nu} = 10 \log \left( \frac{\big[\boldsymbol{J}_D^{-1}|_{\ve{\tilde{x}}_\mathrm{rect}}\big]_{11/22}}{\big[\boldsymbol{J}_D^{-1}|_{\ve{\tilde{x}}}\big]_{11/22}} \right),
\end{equation}
with respect to a rectangular pulse $\ve{\tilde{x}}_\mathrm{rect}$ of bandwidth $B = f_s$, as it is used in Global Navigation Satellite Systems (GNSS). For the following results, the expectation \eqref{eq:channel:sensitivity:matrix} with respect to $p(\boldsymbol{\theta})$ is computed using \textit{Hermite-Gaussian quadrature}.
\vspace{-.35cm}
\setlength{\figurewidth}{0.4\textwidth}
\setlength{\figureheight}{4.7cm}
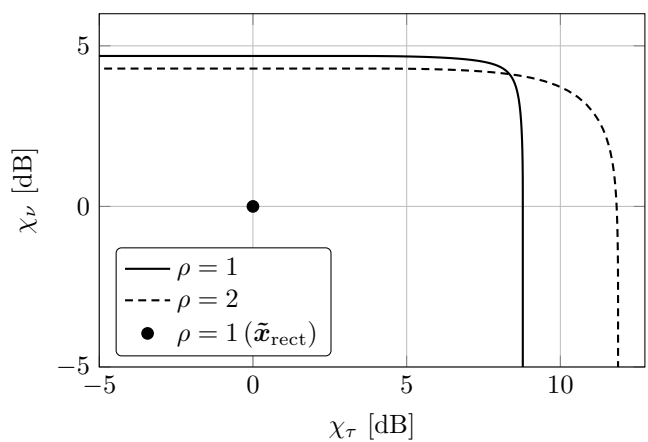
\begin{figure}[!htbp]
	\centering
	\begin{tikzpicture}

\begin{axis}[%
width=\figurewidth,
height=\figureheight,
scale only axis,
ymin=-5,
ymax=6,
xmin=-5,
xmax=12.75,
axis background/.style={fill=white},
xmajorgrids,
ymajorgrids,
ylabel={$\chi_\nu \,\, \mathrm{[dB]}$ },
xlabel={$\chi_\tau  \,\, \mathrm{[dB]}$},
legend pos = south west,
legend cell align=left,
]
\addplot [color=black,solid,thick]
  table[row sep=crcr]{%
-13.0102999566398	4.68467775633226\\
3.42878183214605	4.68467775633226\\
4.04128742203242	4.68229982979945\\
4.63858122579963	4.67637892671706\\
5.19137062344999	4.66653689077827\\
5.67719340656946	4.65307249177975\\
6.08628078376741	4.63686630504978\\
6.42115075195249	4.61900649840257\\
6.6917717058308	4.60043040316801\\
6.91042989648544	4.58176984590955\\
7.08856785669339	4.56337012195506\\
7.23557166755177	4.54537579815448\\
7.35867539867078	4.52781325426345\\
7.46329981224192	4.51064801716568\\
7.55347121933807	4.49381865900323\\
7.6321867281036	4.47725483836333\\
7.70169589875213	4.46088603571822\\
7.76370666945089	4.44464526024342\\
7.81953258225824	4.42847019939693\\
7.87019735854405	4.4123031459138\\
7.9165093420478	4.39609039026266\\
7.95911486662721	4.37978141810199\\
7.99853688866056	4.36332807062192\\
8.03520326414661	4.34668373339183\\
8.06946768876348	4.329802573684\\
8.10162538784506	4.31263882439272\\
8.13192500921203	4.29514610266794\\
8.16057773935192	4.27727674704747\\
8.18776436683971	4.25898115501038\\
8.21364081186675	4.24020710184536\\
8.23834249774339	4.22089902066221\\
8.26198783952645	4.20099722178737\\
8.28468105328314	4.18043702734751\\
8.30651443803126	4.15914779328302\\
8.32757024504606	4.13705178603281\\
8.34792222185313	4.11406287430191\\
8.36763689796883	4.09008498713063\\
8.38677466430108	4.06501027719323\\
8.40539068666111	4.03871691184504\\
8.42353568503573	4.01106639248081\\
8.44125660336488	3.98190027326116\\
8.45859718896355	3.9510361103797\\
8.47559849591233	3.91826241877386\\
8.49229932222076	3.88333233882022\\
8.5087365857818	3.84595561294974\\
8.52494563833613	3.80578832962725\\
8.54096050875077	3.76241969319441\\
8.55681405514434	3.71535479935842\\
8.57253798691172	3.66399200559032\\
8.58816268777611	3.60759294149323\\
8.60371672173331	3.54524245716265\\
8.61922582215233	3.47579481522216\\
8.63471102960173	3.39780120305151\\
8.65018542521682	3.30941235421278\\
8.66564856420958	3.20824936274968\\
8.68107721965693	3.09123741457949\\
8.69641045332656	2.954405380667\\
8.71152667528454	2.7926779821856\\
8.72621130401696	2.59974235561751\\
8.74011832454956	2.36817564030395\\
8.75274103188251	2.0901660910087\\
8.76342667498284	1.75923162602818\\
8.7714808660047	1.3730312732563\\
8.77637270843235	0.936387537898737\\
8.77795574573341	0.462503038039669\\
8.77795574573341	-13.0102999566398\\
};
\addlegendentry{$\rho=1$};
\addplot [color=black,densely dashed,thick]
  table[row sep=crcr]{%
-13.0102999566398	4.29220730566107\\
3.8693247217274	4.29220730566107\\
3.90099911437486	4.29221616052823\\
3.93508663120883	4.29221648920205\\
3.97186313221529	4.29220603130192\\
4.01164633821962	4.29218193229713\\
4.05480366310722	4.29214056776596\\
4.10176174998644	4.29207730825596\\
4.15301811482783	4.29198620193109\\
4.20915539325384	4.29185954234926\\
4.27085879018328	4.29168727417414\\
4.33893743921984	4.29145616795816\\
4.41435046733834	4.29114866258593\\
4.49823858324109	4.29074122475005\\
4.59196186838086	4.29020200008188\\
4.69714396003152	4.28948741695463\\
4.81572162150133	4.28853723224899\\
4.94999615153083	4.28726725281844\\
5.10267807978344	4.28555859879777\\
5.27690735829112	4.283241881935\\
5.47621542140926	4.28007410715392\\
5.70437122262894	4.27570571379779\\
5.96502373432403	4.26963562937696\\
6.26103775474176	4.26115486967434\\
6.59347348706416	4.24928566904645\\
6.96038033258694	4.23273187847327\\
7.35600687106657	4.20985473387984\\
7.77137607166602	4.17864946567348\\
8.19663874454524	4.13662201207111\\
8.62387486069566	4.08045732999926\\
9.04781434760806	4.00558535291546\\
9.46352387146208	3.90610926032358\\
9.86315607348033	3.77571263505829\\
10.234956312194	3.60979816933618\\
10.5659261542631	3.40814359450072\\
10.8464880497073	3.17635078600319\\
11.0737656095423	2.92468905781556\\
11.2514749606148	2.66489748589818\\
11.3873341426377	2.40702026510429\\
11.4901647747338	2.15779377183407\\
11.5680125993157	1.92061263918273\\
11.6273962297696	1.69631249170555\\
11.6732598800183	1.4840890751133\\
11.7092221359188	1.28223259716105\\
11.737885751292	1.08861576327714\\
11.7611108583477	0.900983265259849\\
11.7802264711811	0.717111936368804\\
11.796184458121	0.534899548715226\\
11.8096684352984	0.352422893894474\\
11.8211700738081	0.167990688739545\\
11.8310429192602	-0.0197954051452566\\
11.8395411716507	-0.211969154223536\\
11.8468485778751	-0.409125111467653\\
11.8531007157569	-0.611356965008514\\
11.8584024685049	-0.818229051010994\\
11.8628414330219	-1.02879841019088\\
11.8664974450964	-1.24169440885971\\
11.8694483426035	-1.45524809551964\\
11.8717723790229	-1.66765003037369\\
11.8735480823246	-1.877108890571\\
11.8748525825853	-2.08198575843006\\
11.8757593969296	-2.28088834785283\\
11.8763364112364	-2.4727208481497\\
11.8766444660328	-2.6566943701048\\
11.8767366605723	-2.83230814705983\\
11.8767366605723	-13.0102999566398\\
};
\addlegendentry{$\rho=2$};
\addplot [color=black,only marks,mark=*,mark options={solid}, thick]
table[row sep=crcr]{%
0	0\\
};
\addlegendentry{$\rho=1\, (\ve{\tilde{x}}_{\mathrm{rect}})$};
\end{axis}
\end{tikzpicture}%
	\caption{Pareto regions for bandwidths $B=\rho f_s$ with $f_s = 10 \mathrm{MHz}$}
	\label{fig:Pareto_Different_Bandwidths}
\end{figure}
\subsection{Pareto-Optimal Region - Fixed Sampling Rate}
For a setting where  $T_0 = 10 \mu \mathrm{s}$, $f_s = 10 \mathrm{MHz}$, $\sigma_\nu = 5 \mathrm{kHz}$ and $\sigma_\tau= 10 \mathrm{ns}$, Fig. \ref{fig:Pareto_Different_Bandwidths} shows the Pareto-optimal regions for different bandwidths $B = \rho f_s$. Note that here for all systems the same sampling frequency $f_s$ has been used. The results indicate that a potential performance gain of roughly $12 \mathrm{dB}$ for delay estimation and $4 \mathrm{dB}$  for Doppler estimation can be obtained when optimizing the transmit system for $\rho=2$. Note that when increasing the transceive bandwidth $B$ from $\rho=1$ to $\rho=2$, two main effects affect the estimation performance. First, a larger transmit bandwidth is beneficial for the delay estimation due to high-frequency signal parts. On the other hand, a higher receive filter bandwidth results in a larger noise power at the receiver and therefore in a lower Doppler estimation accuracy. However, the optimized system is able to compensate this effect by efficiently using the available transmit spectrum, which leads to a moderate loss.
\subsection{Pareto-Optimal Region - Fixed Bandwidth}
In the previous section, we have seen that optimized waveforms have the potential to increase the accuracy of delay-Doppler estimation methods. We now investigate the estimation performance for a fixed transmit bandwidth $B = 10 \mathrm{MHz}$, a signal period $T_0 = 10 \mu \mathrm{s}$ and different sampling frequencies $f_s = \frac{B}{\kappa}$. In order to focus on the case with undersampling we consider setups where $\kappa > 1$.
\begin{figure}[!htbp]
	\centering
	\begin{tikzpicture}

\begin{axis}[%
width=\figurewidth,
height=\figureheight,
scale only axis,
ymin=-5,
ymax=6,
xmin=-5,
xmax=12.75,
axis background/.style={fill=white},
xmajorgrids,
ymajorgrids,
ylabel={$\chi_\nu \,\, \mathrm{[dB]}$ },
xlabel={$\chi_\tau  \,\, \mathrm{[dB]}$},
legend pos = south west,
legend cell align=left,
]
\addplot [color=black,solid,thick]
table[row sep=crcr]{%
-13.0102999566398	4.68467775633226\\
3.42878183214605	4.68467775633226\\
4.04128742203242	4.68229982979945\\
4.63858122579963	4.67637892671706\\
5.19137062344999	4.66653689077827\\
5.67719340656946	4.65307249177975\\
6.08628078376741	4.63686630504978\\
6.42115075195249	4.61900649840257\\
6.6917717058308	4.60043040316801\\
6.91042989648544	4.58176984590955\\
7.08856785669339	4.56337012195506\\
7.23557166755177	4.54537579815448\\
7.35867539867078	4.52781325426345\\
7.46329981224192	4.51064801716568\\
7.55347121933807	4.49381865900323\\
7.6321867281036	4.47725483836333\\
7.70169589875213	4.46088603571822\\
7.76370666945089	4.44464526024342\\
7.81953258225824	4.42847019939693\\
7.87019735854405	4.4123031459138\\
7.9165093420478	4.39609039026266\\
7.95911486662721	4.37978141810199\\
7.99853688866056	4.36332807062192\\
8.03520326414661	4.34668373339183\\
8.06946768876348	4.329802573684\\
8.10162538784506	4.31263882439272\\
8.13192500921203	4.29514610266794\\
8.16057773935192	4.27727674704747\\
8.18776436683971	4.25898115501038\\
8.21364081186675	4.24020710184536\\
8.23834249774339	4.22089902066221\\
8.26198783952645	4.20099722178737\\
8.28468105328314	4.18043702734751\\
8.30651443803126	4.15914779328302\\
8.32757024504606	4.13705178603281\\
8.34792222185313	4.11406287430191\\
8.36763689796883	4.09008498713063\\
8.38677466430108	4.06501027719323\\
8.40539068666111	4.03871691184504\\
8.42353568503573	4.01106639248081\\
8.44125660336488	3.98190027326116\\
8.45859718896355	3.9510361103797\\
8.47559849591233	3.91826241877386\\
8.49229932222076	3.88333233882022\\
8.5087365857818	3.84595561294974\\
8.52494563833613	3.80578832962725\\
8.54096050875077	3.76241969319441\\
8.55681405514434	3.71535479935842\\
8.57253798691172	3.66399200559032\\
8.58816268777611	3.60759294149323\\
8.60371672173331	3.54524245716265\\
8.61922582215233	3.47579481522216\\
8.63471102960173	3.39780120305151\\
8.65018542521682	3.30941235421278\\
8.66564856420958	3.20824936274968\\
8.68107721965693	3.09123741457949\\
8.69641045332656	2.954405380667\\
8.71152667528454	2.7926779821856\\
8.72621130401696	2.59974235561751\\
8.74011832454956	2.36817564030395\\
8.75274103188251	2.0901660910087\\
8.76342667498284	1.75923162602818\\
8.7714808660047	1.3730312732563\\
8.77637270843235	0.936387537898737\\
8.77795574573341	0.462503038039669\\
8.77795574573341	-13.0102999566398\\
};
\addlegendentry{$\kappa=1$};
\addplot [color=black,densely dashed,thick]
table[row sep=crcr]{%
-13.0102999566398	4.57957618958201\\
-2.88366350301073	4.57957618958201\\
-2.76367464134824	4.57959992737693\\
-2.63863096433874	4.57948111608594\\
-2.50830910710773	4.57919360979315\\
-2.37248822665369	4.57870688265093\\
-2.23095406048749	4.57798532902506\\
-2.08350400706313	4.57698747321311\\
-1.92995337628471	4.57566508794883\\
-1.77014294434893	4.57396222666552\\
-1.60394791172304	4.57181418293936\\
-1.43128829702288	4.56914640222549\\
-1.25214069257349	4.56587338628102\\
-1.06655114924612	4.56189764938599\\
-0.874648741342712	4.55710880656085\\
-0.676659086684497	4.55138289492643\\
-0.472916775970168	4.54458204566481\\
-0.263875332996336	4.5365546289156\\
-0.0501130447552543	4.52713597856964\\
0.167667142388905	4.51614975891938\\
0.388645357578573	4.50340995357845\\
0.611901407225976	4.48872333914521\\
0.836440603871788	4.47189216409334\\
1.06122792794315	4.45271661503532\\
1.28522816409989	4.43099655884875\\
1.50744828190278	4.40653204468114\\
1.72697742346985	4.37912216579664\\
1.94301974338552	4.34856211795799\\
2.15491616074121	4.31463860915428\\
2.36215270554039	4.2771241034849\\
2.56435520043313	4.23577064003821\\
2.76127202089502	4.19030409684732\\
2.95274818205758	4.14041975280086\\
3.13869475970091	4.08577986115535\\
3.31905765530912	4.0260137367104\\
3.49378913658703	3.96072062568408\\
3.6628246852818	3.88947540610925\\
3.82606670673532	3.81183696581974\\
3.98337576603449	3.72735891367172\\
4.13456928871208	3.63560207822152\\
4.27942710408771	3.5361480218335\\
4.4177027851965	3.42861254635725\\
4.54913941612215	3.31265790644353\\
4.67348817822205	3.188002211269\\
4.79052799806627	3.05442432458008\\
4.90008446650117	2.91176249862941\\
5.00204634737131	2.75990500768807\\
5.0963782582736	2.59877115085258\\
5.18312850736356	2.42828109612087\\
5.26243155868135	2.24831302260364\\
5.3345050945096	2.05864575052648\\
5.39964205383498	1.85888441710034\\
5.45819826118715	1.64836575071431\\
5.51057624781535	1.42603837826638\\
5.55720556582139	1.19031328224544\\
5.5985193226628	0.938882385477125\\
5.63492598426755	0.668514714593428\\
5.66677526619573	0.374870488582405\\
5.69431857201318	0.0524407822854734\\
5.71767063531649	-0.305163096203599\\
5.73679238429337	-0.704216299550832\\
5.75153121744482	-1.14885940751261\\
5.76175094553465	-1.6382476704078\\
5.76752703707003	-2.16416395152494\\
5.7692943555743	-2.71110689570292\\
5.7692943555743	-13.0102999566398\\
};
\addlegendentry{$\kappa=2$};
\addplot [color=black,loosely dashed,thick]
table[row sep=crcr]{%
-13.0102999566398	4.53762964207443\\
-7.88140225537725	4.53762964207443\\
-7.3900736667784	4.53618093859101\\
-6.82669299894257	4.53010768715797\\
-6.1980815184661	4.51698025579058\\
-5.52014849153453	4.49397552586083\\
-4.81771433402378	4.45841445316383\\
-4.12128321920248	4.40862518358906\\
-3.4609525237582	4.34478957788332\\
-2.85914908334414	4.26911486955454\\
-2.32411253868511	4.18466712406328\\
-1.84381610335354	4.09236813274537\\
-1.37544389883245	3.98489142100718\\
-0.820732557188867	3.83224879680107\\
0.00556294407328907	3.54458644220496\\
1.26266611170244	2.91434975221219\\
2.63652432079052	1.75574754436661\\
3.5761523371427	0.360354373489942\\
4.0692421506071	-0.806961499948585\\
4.31850007200543	-1.61867981029804\\
4.44930926304319	-2.13940915744826\\
4.52024205987431	-2.45844975640347\\
4.55906352597683	-2.64613990815306\\
4.57998916952882	-2.75157795091977\\
4.59084742698636	-2.80754105982923\\
4.59615078765072	-2.83519552920078\\
4.59852896018417	-2.84766613329981\\
4.59947783125043	-2.85265391846611\\
4.59979998525159	-2.854348912945\\
4.59988670102678	-2.85480529964593\\
4.59990302501925	-2.8548912196031\\
4.59990468796757	-2.85489997250664\\
4.5999047358382	-2.85490022447329\\
4.59990473588204	-2.85490022470409\\
4.59990478375264	-2.8549004766706\\
4.59990644669592	-2.85490922957021\\
4.59992277012113	-2.8549951490284\\
4.60000946716221	-2.8554515192569\\
4.60033131998572	-2.85714624891435\\
4.60127718864567	-2.86213139411002\\
4.60363404709652	-2.87458325729846\\
4.60881994344115	-2.90213445102539\\
4.61914596210564	-2.95762977281628\\
4.63800778858943	-3.06125642227171\\
4.66977747512379	-3.24278724575656\\
4.71899713289124	-3.54315062964412\\
4.78844783300165	-4.01338044944205\\
4.87624834443138	-4.70709181043441\\
4.9737406624467	-5.66164526418571\\
};
\addlegendentry{$\kappa=4$};
\addplot [color=black,only marks,mark=*,mark options={solid}, thick]
table[row sep=crcr]{%
	0	0\\
};	
\addlegendentry{$\kappa=1\, (\ve{\tilde{x}}_{\mathrm{rect}})$};
\addplot [color=black,only marks,mark=*,mark options={scale=1.1, solid, fill=white}, thick]
table[row sep=crcr]{%
	4.5491 3.3127\\
};	

\node[right] at (axis cs:4.9,3.55){$\ve{\tilde{x}}^\star$};

\end{axis}
\end{tikzpicture}%
	\caption{Pareto regions for rates $f_s=\frac{B}{\kappa}$ with $B = 10 \mathrm{MHz}$}
	\label{fig:Pareto_Different_Rates}
\end{figure}
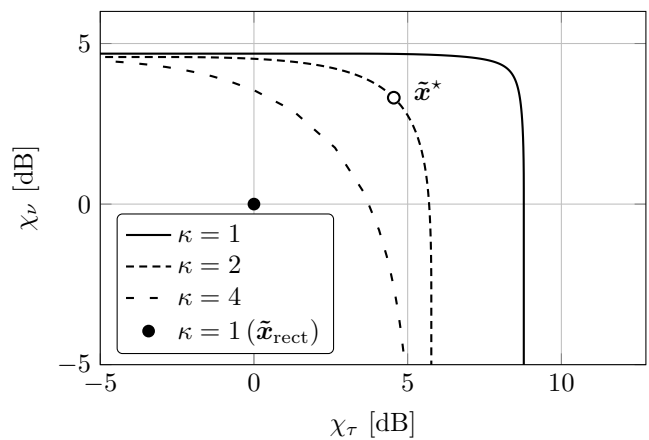
Fig. \ref{fig:Pareto_Different_Rates} shows the Pareto regions of the optimized waveforms with respect to a rectangular signal. Note that the sampling rate for the reference system is held constant, while the sampling rate of the optimized system decreases with increasing $\kappa$. This indicates that although lower sampling rates are used, the optimized waveform design still bears the potential to provide high estimation accuracy.

\subsection{Simulation Results}\label{ss:Simulation_Results}
To verify that the optimization based on the EFIM yields substantial performance gains for practical scenarios, we conduct Monte-Carlo simulations with randomly generated noise $\ve{\eta}$ and channel parameters $\ve{\theta}$. As the channel $\gamma$ is in general not known to the receiver, we use the hybrid maximum likelihood-maximum a posteriori (ML-MAP) estimator \cite[p. 12]{Trees07} 
\begin{align}\label{def:jmap:ml}
\begin{pmatrix}
\hat{\gamma}_{\mathrm{ML}}(\boldsymbol{y})\\
\boldsymbol{\hat{\theta}}_{\mathrm{MAP}}(\boldsymbol{y})
\end{pmatrix}
&= \arg\underset{\boldsymbol{\theta},\gamma}{\max} \, \big(\ln p(\boldsymbol{y}|\boldsymbol{\theta},\gamma) + \ln p(\boldsymbol{\theta})\big).
\end{align}
For simulations we use $T_0 = 10\mu \mathrm{s}$ and $B = 10\mathrm{MHz}$. We compare the MSE of a rectangular pulse signal with $f_s=10\mathrm{MHz}$ and the optimized transmit signal $\ve{\tilde{x}}^\star$ with $f_s = 5 \mathrm{MHz}$, i.e., $\kappa=2$. The transmitter design $\ve{\tilde{x}}^\star$ used for the simulations corresponds to the point of the Pareto-region in Fig. \ref{fig:Pareto_Different_Rates} with largest distance to the origin. Fig. \ref{fig:MSE_Tau_Equal_Weight} and Fig. \ref{fig:MSE_Nu_Equal_Weight} show the normalized empirical mean squared error (NMSE)
\begin{equation}
\mathrm{NMSE}_{\hat{\tau}/\hat{\nu}} = \frac{\mathrm{MSE}_{\hat{\tau}/\hat{\nu}}}{\sigma^2_{\tau/\nu}}
\vspace{-.05cm}
\end{equation}
of the hybrid ML-MAP estimator for both systems, where $\mathrm{MSE}_{\hat{\tau}/\hat{\nu}}$ represents the diagonal elements of \eqref{eq:MSE:matrix}, empirically evaluated based on the results of the estimation algorithm \eqref{def:jmap:ml}. The signal-to-noise ratio (SNR) is given by
\begin{equation}
\mathrm{SNR} = \frac{P}{BN_0}.
\end{equation}
\vspace{-.1cm}
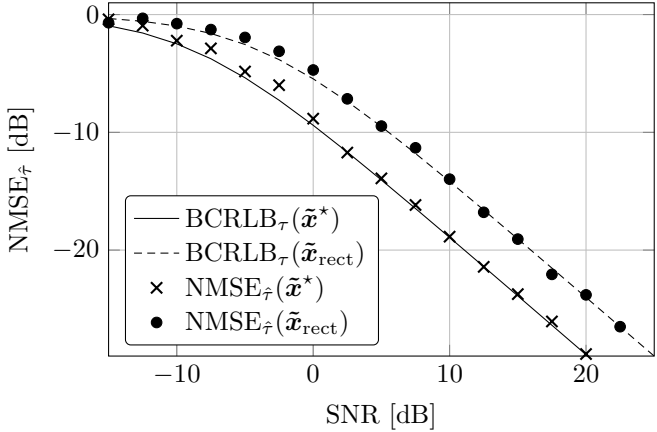
\begin{figure}[h]
	\begin{tikzpicture}

\begin{axis}[%
width=\figurewidth,
height=\figureheight,
at={(0\figurewidth,0\figureheight)},
scale only axis,
xmin=-15,
xmax=25,
xmajorgrids,
ymin=-29,
ymax=1,
ymajorgrids,
axis background/.style={fill=white},
legend cell align=left,
xlabel={$\mathrm{SNR} \,\, \mathrm{[dB]}$},
ylabel={$\mathrm{NMSE}_{\hat{\tau}} \,\, \mathrm{[dB]}$},
scaled y ticks=false,
cycle list name=black white,
legend pos=south west,
]
\addplot [solid]
table[row sep=crcr]{%
-15	-0.947973917582718\\
-12.5	-1.56483435140501\\
-10	-2.48313523591258\\
-7.5	-3.75067409272672\\
-5	-5.36475055955058\\
-2.5	-7.2736443052085\\
-0	-9.40210848373573\\
2.5	-11.6783126927945\\
5	-14.0472047927137\\
7.5	-16.4717006773638\\
10	-18.9286577544366\\
12.5	-21.4042642139691\\
15	-23.8904862884958\\
17.5	-26.3827191468242\\
20	-28.8783452520753\\
22.5	-31.375883693815\\
25	-33.8744988447353\\
};
\addlegendentry{$\mathrm{BCRLB}_{\tau}(\ve{\tilde{x}}^\star)$};
\addplot [densely dashed]
table[row sep=crcr]{%
-15	-0.331954169144444\\
-12.5	-0.573818777130086\\
-10	-0.973222396862178\\
-7.5	-1.60372703780636\\
-5	-2.53900669388154\\
-2.5	-3.82472464927109\\
-0	-5.45538489273941\\
2.5	-7.3773378458128\\
5	-9.51494490959796\\
7.5	-11.7970358644904\\
10	-14.169516352988\\
12.5	-16.5961270923002\\
15	-19.0543058889181\\
17.5	-21.5306099718848\\
20	-24.017227763333\\
22.5	-26.509684240703\\
25	-29.0054364432437\\
};
\addlegendentry{$\mathrm{BCRLB}_{\tau} (\ve{\tilde{x}}_\mathrm{rect})$};
\addplot [only marks, mark=x, thick,mark options={scale=1.5}]
table[row sep=crcr]{%
-15	-0.403933960569669\\
-12.5	-0.937376799519823\\
-10	-2.21550695286609\\
-7.5	-2.87163521304082\\
-5	-4.83457539710767\\
-2.5	-5.99526855455307\\
-0	-8.83757982973824\\
2.5	-11.7198056378713\\
5	-13.9107257717091\\
7.5	-16.1716732715895\\
10	-18.8592624926874\\
12.5	-21.4236232338314\\
15	-23.7395374267311\\
17.5	-26.062236718159\\
20	-28.8278724950411\\
22.5	-31.4333448204117\\
25	-33.6881161654101\\
};
\addlegendentry{$\mathrm{NMSE}_{\hat{\tau}}(\ve{\tilde{x}}^\star)$};
\addplot [only marks, mark = *]
table[row sep=crcr]{%
-15	-0.693795535604502\\
-12.5	-0.317173579359192\\
-10	-0.770261233243398\\
-7.5	-1.274159237207\\
-5	-1.94098903035029\\
-2.5	-3.11811986401794\\
-0	-4.70564146946555\\
2.5	-7.15589405234193\\
5	-9.46987204563393\\
7.5	-11.3156436207543\\
10	-13.9764490298791\\
12.5	-16.7909268272837\\
15	-19.0683094427502\\
17.5	-22.0633734253195\\
20	-23.8030054438259\\
22.5	-26.5065667834306\\
25	-29.0814342183174\\
};
\addlegendentry{$\mathrm{NMSE}_{\hat{\tau}}(\ve{\tilde{x}}_\mathrm{rect})$};
\end{axis}
\end{tikzpicture}%
	\caption{MSE and BCRLB - Time-delay $\tau$}
	\label{fig:MSE_Tau_Equal_Weight}
	\vspace{-.3cm}
\end{figure}
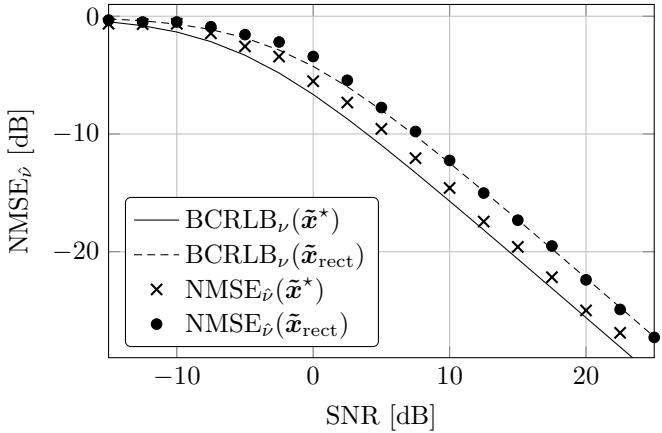
\begin{figure}[h]
	\begin{tikzpicture}

\begin{axis}[%
width=\figurewidth,
height=\figureheight,
at={(0\figurewidth,0\figureheight)},
scale only axis,
xmin=-15,
xmax=25,
xmajorgrids,
ymin=-29,
ymax=1,
ymajorgrids,
axis background/.style={fill=white},
legend cell align=left,
xlabel={$\mathrm{SNR} \,\, \mathrm{[dB]}$},
ylabel={$\mathrm{NMSE}_{\hat{\nu}} \,\, \mathrm{[dB]}$},
cycle list name=black white,
legend pos=south west,
]
\addplot [solid]
table[row sep=crcr]{%
-15	-0.470524439968727\\
-12.5	-0.804413574355385\\
-10	-1.34131324404043\\
-7.5	-2.15764862099418\\
-5	-3.31283501528766\\
-2.5	-4.82139633146154\\
-0	-6.645067049821\\
2.5	-8.71270011997996\\
5	-10.9492139127694\\
7.5	-13.2937065394002\\
10	-15.7037485582687\\
12.5	-18.1523307306516\\
15	-20.6231466820186\\
17.5	-23.1066487238315\\
20	-25.5973436348595\\
22.5	-28.0921022275238\\
25	-30.5891519863963\\
};
\addlegendentry{$\mathrm{BCRLB}_{\nu} (\ve{\tilde{x}}^\star)$};
\addplot [densely dashed]
table[row sep=crcr]{%
-15	-0.222881742995533\\
-12.5	-0.388761436626724\\
-10	-0.668944058120939\\
-7.5	-1.12648453972826\\
-5	-1.83725628990292\\
-2.5	-2.8700499957355\\
-0	-4.2573950640174\\
2.5	-5.97837377737191\\
5	-7.96991294283819\\
7.5	-10.1554594903404\\
10	-12.468085670317\\
12.5	-14.8590543055494\\
15	-17.2965172365063\\
17.5	-19.7609503984968\\
20	-22.2408209955364\\
22.5	-24.7294602805463\\
25	-27.223058601909\\
};
\addlegendentry{$\mathrm{BCRLB}_{\nu} (\ve{\tilde{x}}_\mathrm{rect})$};
\addplot [only marks, mark=x, thick,mark options={scale=1.5}]
table[row sep=crcr]{%
-15	-0.634782750861797\\
-12.5	-0.668606264670528\\
-10	-0.643769294698706\\
-7.5	-1.45065443237508\\
-5	-2.5653562029328\\
-2.5	-3.42526417344535\\
-0	-5.52086764312785\\
2.5	-7.33079623718833\\
5	-9.56806298399493\\
7.5	-12.052798552045\\
10	-14.5851721134425\\
12.5	-17.4383624474844\\
15	-19.5902523235028\\
17.5	-22.1643248722267\\
20	-24.9837078389492\\
22.5	-26.8867092062875\\
25	-29.4543889254823\\
};
\addlegendentry{$\mathrm{NMSE}_{\hat{\nu}}(\ve{\tilde{x}}^\star)$};
\addplot [only marks, mark = *]
table[row sep=crcr]{%
-15	-0.327974043777389\\
-12.5	-0.493413144138805\\
-10	-0.491203195523659\\
-7.5	-0.894391320377133\\
-5	-1.56170466144567\\
-2.5	-2.18907675835425\\
-0	-3.42082257851574\\
2.5	-5.43982291056628\\
5	-7.75310383750146\\
7.5	-9.78833350647525\\
10	-12.2480367182927\\
12.5	-15.0179997294383\\
15	-17.3206366370425\\
17.5	-19.5152947022775\\
20	-22.3864214860626\\
22.5	-24.901714324417\\
25	-27.2750447420843\\
};
\addlegendentry{$\mathrm{NMSE}_{\hat{\nu}}(\ve{\tilde{x}}_\mathrm{rect})$};
\end{axis}
\end{tikzpicture}%
	\caption{MSE and BCRLB - Doppler-shift $\nu$}
	\label{fig:MSE_Nu_Equal_Weight}
	\vspace{-.4cm}
\end{figure}
\vspace{-.0cm}
It is observed that for low SNR the MSE saturates at $\sigma_{\tau, \nu}^2$, since in this case the estimation merely relies on the prior information $p(\ve{\theta})$. In the high SNR regime, the MSE of the hybrid ML-MAP estimator shows close correspondence with the BCRLB and the estimator benefits from the waveform optimization. For moderate to high SNR values the performance gain is roughly $4.5\mathrm{dB}$ for the estimation of the time-delay and $3.5\mathrm{dB}$ for the Doppler-shift estimation. This corresponds to the findings from the Pareto-region in Fig. \ref{fig:Pareto_Different_Rates}.
\section{Conclusion}
\label{sec:conclusion}
We have derived an optimization framework for the transmit waveform of an undersampled pilot-based channel estimation system. By employing the BCRLB, the transmitter design problem was reformulated as a maximization problem with respect to the expected Fisher information matrix. A frequency domain representation of the receive signal allows one to find an analytical solution to the maximization problem via an Eigenvalue decomposition. The BCRLB of the optimized waveforms can be used to approximately characterize the Pareto-optimal design region with respect to other delay-Doppler estimation methods. Further, our results show that using optimized transmit waveforms enables the receiver to operate significantly below the Nyquist sampling rate while maintaining high delay-Doppler estimation accuracy. Finally, Monte-Carlo simulations support the practical impact of the considered transmit design problem.
\vspace{-.1cm}
\bibliographystyle{IEEEtran}

\begin{thebibliography}{10}
\providecommand{\url}[1]{#1}
\csname url@samestyle\endcsname
\providecommand{\newblock}{\relax}
\providecommand{\bibinfo}[2]{#2}
\providecommand{\BIBentrySTDinterwordspacing}{\spaceskip=0pt\relax}
\providecommand{\BIBentryALTinterwordstretchfactor}{4}
\providecommand{\BIBentryALTinterwordspacing}{\spaceskip=\fontdimen2\font plus
\BIBentryALTinterwordstretchfactor\fontdimen3\font minus
  \fontdimen4\font\relax}
\providecommand{\BIBforeignlanguage}[2]{{%
\expandafter\ifx\csname l@#1\endcsname\relax
\typeout{** WARNING: IEEEtran.bst: No hyphenation pattern has been}%
\typeout{** loaded for the language `#1'. Using the pattern for}%
\typeout{** the default language instead.}%
\else
\language=\csname l@#1\endcsname
\fi
#2}}
\providecommand{\BIBdecl}{\relax}
\BIBdecl

\bibitem{Sadler06}
B.~Sadler and R.~Kozick, ``A survey of time delay estimation performance bounds,'' in \emph{Fourth IEEE Workshop Sensor Array and Multichannel Processing}, Waltham, MA, 2006, pp. 282-288.

\bibitem{Verhelst15}
M. Verhelst and A. Bahai, ``Where analog meets digital: Analog-to-information conversion and beyond,'' \emph{IEEE Solid State Circuits Mag.}, vol. 7, no. 3, pp. 67-80, Sep. 2015.

\bibitem{Jakobsson98}
A.~Jakobsson, A.~L. Swindlehurst, and P.~Stoica, ``Subspace-based estimation of time delays and {D}oppler shifts,'' \emph{IEEE Trans. Signal Process.}, vol.~46, no.~9, pp. 2472--2483, Sep. 1998.

\bibitem{Jakobsson98_2}
A.~Jakobsson and A.~L.~Swindlehurst ``A time domain method for joint estimation of time delays, Doppler shifts and spatial signatures,'' in \emph{Ninth IEEE Signal Processing Workshop on Statistical Signal and Array Processing}, Portland, OR, 1998, pp. 388-391.

\bibitem{Friedlander84}
B.~Friedlander, ``On the Cram\'{e}r-Rao bound for time delay and Doppler estimation,'' \emph{IEEE Trans. Inf. Theory}, vol. 30, no. 3, pp. 575-580, May 1984.

\bibitem{Antreich11}
F.~Antreich, ``Array processing and signal design for timing
	synchronization,'' Ph.D. dissertation, NWS, 
TUM, M\"{u}nchen, 2011.

\bibitem{Jin95}
Q.~Jin, K.~Wong and Z.~Luo, ``The estimation of time delay and Doppler stretch of wideband signals,'' \emph{IEEE Trans. Sig, Proc.}, vol. 43, no. 4, pp. 904-916, Apr. 1995.

\bibitem{Stein14}
M.~Stein, A.~Lenz, A.~Mezghani, and J.~Nossek, ``Optimum analog receive filters for detection and inference under a sampling rate constraint,'' in \emph{IEEE Int. Conf. Acoustics, Speech and Signal Processing (ICASSP)}, Florence, 2014, pp. 1827--1831.

\bibitem{Lenz15}
A.~Lenz, M.~Stein, and J.~A. Nossek, ``Signal parameter estimation performance under a sampling rate constraint,'' in \emph{49th Asilomar Conf. Signals, Systems and Computers}, Pacific Grove, CA, 2015, pp. 503--507.
  
\bibitem{Khayambashi14}
M.~Khayambashi and A.~L.~Swindlehurst, ``Filter design for a compressive sensing delay and {D}oppler estimation framework,'' in \emph{48th Asilomar Conf. Signals, Systems and Computers}, Pacific Grove, CA, 2014, pp. 627--631.

\bibitem{Trees07}
H. L. Van Trees and K. L. Bell, \emph{Bayesian Bounds for Parameter Estimation and Nonlinear Filtering/Tracking.} Piscataway, NJ: Wiley-IEEE Press, 2007.

\bibitem{Stein14_2}
M.~Stein, M.~Castaneda, A.~Mezghani, and J.~Nossek, ``Information-preserving transformations for signal parameter estimation,'' \emph{IEEE Signal Process. Lett.}, vol.~21, no.~7, pp. 866--870, July 2014.

\end{thebibliography}

\end{document}